\documentclass[aps,prd,preprint,showpacs,showkeys,preprintnumbers,amsmath,amssymb]{revtex4-1}
\usepackage{graphicx}

\begin{document}

\title{On Non-commutative Geodesic Motion}

\author{S. C. Ulhoa }
\email{sc.ulhoa@gmail.com}
\author{R. G. G. Amorim}
\email{ronniamorim@gmail.com} \affiliation{Instituto de F\'{i}sica,
Universidade de Bras\'{i}lia, 70910-900, Bras\'{i}lia, DF,
Brazil.\\Faculdade Gama, Universidade de Bras\'{i}lia, Setor Leste
(Gama), 72444-240, Bras\'{i}lia-DF, Brazil.}

\author{A. F. Santos}
\email{alesandroferreira@fisica.ufmt.br }
\address{Instituto de F\'isica, Universidade Federal de Mato Grosso,
78060-900, Cuiab\'a, MT, Brazil.}

\date{\today}

\begin{abstract}
In this work we study the geodesic motion on a noncommutative space-time. As a result we find a non-commutative geodesic equation and then we derive corrections of the deviation angle per revolution in
terms of the non-commutative parameter when we specify the problem of Mercury's perihelion. In this way, we estimate the noncommutative parameter based in experimental data.
\end{abstract}


\maketitle

\section{Introduction}

In 1845 the french astronomer Le Verrier observed that the perihelion of the planet Mercury precess  faster rate than can be accounted the Newtonian mechanics with the distribution of masses of the solar system well-known until then. The calculations from  Newtonian mechanics present a discrepancy of $43'11'' \pm 0,45''$ in comparison with experimental data. This discovery began different lines of investigation to explain the new phenomena. One of the explanations was the existence of a new planet that would explain the anomaly in Mercury's orbit within the context of Newton's laws \cite{mp1, mp2}. From then the problem of Mercury perihelion became a topic of a hard discussions in the scientific community. However, with the advent of the General Relativity Theory the problem was solved and is considered one of triumphs of the new theory. Recently, questions arising in the study of quantum gravity regained interest in reconsidering these issues in the framework of non-commutative space-time. In this sense, we can cite recent studies that  investigated the modifications introduced by a generalized uncertainly principle in classical orbit of particle \cite{mp3, mp4, mp5, mp6}. The main consequence of these researches is impose a constraint on the minimal observable length and non-commutativity parameter in comparison with observational data of Mercury. In this work our aim is write the geodesic equation that arises from the metric tensor which have been corrected using non-commutative product between tetrads. In addition, we hope to obtain
corrections in terms of the non-commutative of the deviation angle per revolution.

In this way, non-commutative geometry has its origin in the Weyl and
Moyal works, studying quantization procedures in phase
space~\cite{weyl}. Snyder \cite{snyder1, snyder2} was the first to
develop a consistent theory for non-commutative space coordinates,
which was based on representations of Lie algebras. Over the last
decades there is a revival of non-commutative physics, motivated by
some results coming from gravity \cite{kalau, kastler, connes1},
standard model \cite{connes2, varilly1, varilly2}, string
theory~\cite{Szabo2003207,Seiberg:1999vs} and in the understanding
of the quantum Hall effect \cite{belissard}. One particular interest
in this context is the development of representation theories for
non-commutative fields~\cite{PhysRevD.69.045013}. So, as direct
consequence of the non-commutative feature of the operators
correspondent to space-time coordinates, is the impossibility to
precisely measure a particle position. From the mathematical
viewpoint, the simplest algebra of the operators
$\widehat{x}^{\mu}$, that represents the hermitian operators
correspondent to space-time coordinates, is given on anti-symmetric
tensor constant $\alpha^{\mu\nu}$,
\begin{equation}
[\widehat{x}^{\mu},\widehat{ x}^{\nu}]=i\alpha^{\mu\nu}.
\end{equation}
The relations above imply in these uncertain relations
\begin{equation}
\Delta \widehat{x}^{\mu}\Delta \widehat{x}^{\nu}\geq
\frac{1}{2}|\alpha^{\mu\nu}|.
\end{equation}
These last relations suggest that, at distances of
$\sqrt{|\alpha^{\mu\nu}|}$ order, effects of the non-commutative
space-time turns out to be relevant, showing the end of the
classical model of space-time and the beginning of a new geometric
structure. Usually the non-commutativity is introduced by means the
use of the Moyal product~\cite{snyder1} defined as

\begin{small}
\begin{eqnarray}
f(x) \star g(x) &\equiv& \exp \left(  {i \over 2} \alpha^{\mu\nu}
{\partial \over \partial x^\mu}
{\partial \over \partial y^\nu} \right) f(x) g(y) |_{y \rightarrow x} \nonumber\\
&=& f(x) g(x) + {i\over 2} \alpha^{\mu \nu}
\partial_\mu f  \partial_\nu  g + {1 \over 2!}  {\left( i \over 2 \right)^2} \alpha^{\mu_1\nu_1} \alpha^{\mu_2\nu_2}(\partial_{\mu_1}   \partial_{\mu_2} f )(\partial_{\nu_1}   \partial_{\nu_2} g )+ \cdots\label{starproduct}\nonumber\\
\end{eqnarray}
\end{small}
where $\alpha^{\mu\nu}$ is the non-commutative parameter. This is
accomplished replacing the usual product, in the classical
lagrangian density, by the Moyal product. Such a product is defined
in an arbitrary coordinates system $x^{\mu}$, here the quantity
$y^{\mu}$ is just an auxiliary variable. The article is organized as
follows. In section \ref{ncmc} we present the noncommutative
corrections to metric field. In section \ref{ncge} we calculate the
noncommutative geodesic equation and we present an estimative for
noncommutative parameter. Finally in the last section we present our
concluding remarks. We use natural units such that $G=c=1$.

\section{Non-commutative corrections of metric tensor}\label{ncmc}

Recently it was shown how to obtain the corrections on metric tensor
components due to the non-commutativity of tetrad
field~\cite{Aschieri:2005yw}. We also have used such procedure to
calculate such corrections for Schwarzschild spacetime~\cite{UR}
which can be described by the following line element

\begin{equation}
ds^2=-\left(1-\frac{2M}{r}\right)dt^2+\left(1-\frac{2M}{r}\right)^{-1}dr^2+r^2\left(d\theta^2+\sin^2{\theta}d\phi^2\right)\,.
\end{equation}
From the tetrad formulation of General Relativity we have the
following relation $$g^{\mu\nu}=e^{a\mu}e_{a}\,^{\nu}\,,$$ where
$e^{a\mu}$ is the tetrad or vierbein components. The tetrad field is
related to the reference frame, thus using such a concept and the
above relation, it is possible to completely determine the tetrad
field. Then we use

\begin{small}
\begin{equation}
e^a\,_\mu=\left[ \begin {array}{cccc} \sqrt {-g_{{00}}}&0&0&0
\\\noalign{\medskip}0&\sqrt {g_{{11}}}\sin\theta\cos\phi &\sqrt
{g_{{22}}}\cos\theta\cos\phi &-\sqrt {g_{{33}}
}\sin\phi\\\noalign{\medskip}0&\sqrt {g_{{11}} } \sin\theta\sin\phi
&\sqrt {g_{{22}}}\cos\theta\sin\phi &\sqrt {g_{{33}} }\cos\phi
\\\noalign{\medskip}0&\sqrt {g_{{11}} } \cos\theta &-\sqrt
{g_{{22}}}\sin\theta &0\end {array} \right]\,,
\end{equation}
\end{small}
this tetrad field is adapted to a stationary observer at spatial
infinity.

In order to introduce the Moyal product into the metric tensor, we
follow the approach in ref.~\cite{Aschieri:2005yw}, this will induce
the settlement of a new metric tensor
$g_{\mu\nu}\rightarrow\tilde{g}_{\mu\nu}$. Such a non-commutative
metric tensor reads

$$\tilde{g}_{\mu\nu}=\frac{1}{2}\left(e^a\,_\mu\star
e_{a\nu}+e^a\,_\nu\star e_{a\mu}\right)\,,$$ where the Moyal product
is defined by the relation (\ref{starproduct}). It is important to
note that the non-commutative metric tensor preserves the SO(3,1)
symmetry. Let us consider a change in the reference frame which is
implemented by the following transformation in the tetrad field
$e'^a\,_\mu=\Lambda^a\,_b e^b\,_\mu$, where $\Lambda^a\,_b$ is the
Lorentz matrix. Since $\Lambda^a\,_b$ do not depend on the
coordinates, it is not affected by the Moyal product. As a
consequence the quantity
$\tilde{\eta}^{ab}=\frac{1}{2}\left(e^a\,_\mu\star
e^{b\mu}+e^b\,_\mu\star e^{a\mu}\right)$ behaves like a tensor under
SO(3,1) transformations. However the same thing is not true for the
Diffeomorphism group, yet it is possible to get a tensorial behavior
for $\tilde{g}_{\mu\nu}$. To this purpose it is introduced a
deformed diffeomorphism group which is obtained by a suitable
mapping of the original differential manifold. Thus, by the
appropriate choice of the representation of such functions on the
deformed manifold, $\tilde{g}_{\mu\nu}$ transforms like a
tensor~\cite{Aschieri:2005yw}. It has been showed that corrections
in the metric tensor appears up to second order in the
non-commutative parameter
$\alpha^{\mu\nu}$~\cite{Mukherjee:2006nd,Chaichian2008573}.

We intent to analyze a geodesic movement over a plane
$\theta=\frac{\pi}{2}$, thus the new metric tensor
$\tilde{g}_{\mu\nu}$ will assume a much simpler form, which reads

\begin{eqnarray}
\tilde{g}_{00}&=&-\left(1-\frac{2M}{r}\right)\,,\nonumber\\
\tilde{g}_{11}&=&\left(1-\frac{2M}{r}\right)^{-1}\left[1+\left(1-\frac{2M}{r}\right)^{-2}\left(\frac{\alpha^2M}{2r^3}\right)\right]\,,\nonumber\\
\tilde{g}_{22}&=&r^2\,,\nonumber\\
\tilde{g}_{33}&=&r^2+\frac{\alpha^2}{8}\,,\label{metricacorrigida}
\end{eqnarray}
where $\alpha=\alpha^{13}$. We also have considered that the only
non-vanishing component of the non-commutative parameter is
$\alpha^{13}$.

\section{Non-commutative geodesic equation}\label{ncge}

Let us consider a system composed by a central mass distribution $M$
and a point particle of mass $m$, as stated before we assume that
the movement takes place in a plane $\theta=\frac{\pi}{2}$. Then the
Hamilton-Jacobi equation is
$$\tilde{g}^{\mu\nu}\frac{\partial S}{\partial x^\mu}\frac{\partial
S}{\partial x^\nu}+m^2=0\,,$$ which, after using the corrected
metric tensor, reads
\begin{eqnarray}
&&-\left(1-\frac{2M}{r}\right)^{-1}\left(\frac{\partial S}{\partial t}\right)^2+\left(1-\frac{2M}{r}\right)\left[1+\left(1-\frac{2M}{r}\right)^{-2}\left(\frac{\alpha^2M}{2r^3}\right)\right]^{-1}\left(\frac{\partial S}{\partial r}\right)^2+\nonumber\\
&&+\left(r^2+\frac{\alpha^2}{8}\right)^{-1}\left(\frac{\partial
S}{\partial \phi}\right)^2=-m^2\,.
\end{eqnarray}
Therefore we assume an action given by $S=-E_0 t+L\phi+S_r(r)$,
where $E_0$ and $L$ are the energy and angular momentum
respectively. Then the function $S_r(r)$ is given by

\begin{equation}
 S_r=\int
\left[1+\left(1-\frac{2M}{r}\right)^{-2}\left(\frac{\alpha^2M}{2r^3}\right)\right]^{1/2}\left\{\frac{E_0^2}
{\left(1-\frac{2M}{r}\right)^2}-\frac{\left[m^2+\frac{L^2/r^2}{\left(1+\alpha^2/8r^2\right)}\right]}
{\left(1-\frac{2M}{r}\right)}\right\}^{1/2}dr\,.
\end{equation}
Thus it is possible to obtain equations for $t$ and $\phi$ by means
the action above, $-t+\frac{\partial S_r}{\partial E_0}=const.$ and
$\phi+\frac{\partial S_r}{\partial L}=const.$. Then, after some
algebraic manipulations it yields

\begin{equation}
t=\frac{E_0}{m}\int\frac{\left[1+\left(1-\frac{2M}{r}\right)^{-2}\left(\frac{\alpha^2M}{2r^3}\right)\right]^{1/2}}
{\left(1-\frac{2M}{r}\right)\left\{\frac{E_0^2}{m^2}-\left(1-\frac{2M}{r}\right)\left[1+\frac{\left(\frac{L^2}{m^2r^2}\right)}
{\left(1+\frac{\alpha^2}{8r^2}\right)}\right]\right\}^{1/2}}\,dr\label{tr}
\end{equation}
and

\begin{equation}
\phi=\int\frac{\left(\frac{L}{mr^2}\right)\left[1+\left(1-\frac{2M}{r}\right)^{-2}\left(\frac{\alpha^2M}{2r^3}\right)\right]^{1/2}}
{\left(1+\frac{\alpha^2}{8r^2}\right)\left\{\frac{E_0^2}{m^2}-\left(1-\frac{2M}{r}\right)\left[1+\frac{\left(\frac{L^2}{m^2r^2}\right)}
{\left(1+\frac{\alpha^2}{8r^2}\right)}\right]\right\}^{1/2}}
\,dr\,.\label{phir}
\end{equation}
From equation (\ref{tr}) is possible to get the conditions that lead
to a circular orbit of $m$ around $M$ while formally the relation
(\ref{phir}) defines the orbit itself. We would like to show the
trajectory equation in a more familiar form. For such a purpose we
change variables to $U=1/r$ in (\ref{phir}), where $U=U(\phi)$, and
take the derivative with respect to $\phi$. Hence we obtain

\begin{equation}
(U^{\prime})^2=\left(1+\frac{\alpha^2U^2}{4}\right)\left\{\left[\frac{E_0^2-m^2\left(1-2MU\right)}{L^2}\right]-U^2\left(1-2MU\right)\right\}\left[1-\frac{\alpha^2MU^3}{2\left(1-2MU\right)}\right]\,,
\end{equation}
with $U^\prime=\frac{dU}{d\phi}$. If we again perform a derivative
of above equation with respect to $\phi$ and use $\alpha U<<1$, then
it yields

\begin{equation}
U^{\prime\prime}+U=\frac{m^2M}{L^2}+3MU^2+\alpha^2g(U)\,,\label{t}
\end{equation}
where $g(U)$ is given by
\begin{small}
\begin{eqnarray}
g(U)&=&\left[\frac{E_0^2U+m^2MU^2-m^2U\left(1-2MU\right)}{4L^2}\right]-\frac{MU^3}{2\left(1-2MU\right)^2}\left[\frac{m^2M}{L^2}-U\left(1-3MU\right)\right]-\nonumber\\
&-&\frac{U^3}{8}\left(2-5MU\right)-MU^2\left[\frac{3-2MU}{4\left(1-2MU\right)^3}\right]\left\{\left[\frac{E_0^2-m^2\left(1-2MU\right)}{L^2}\right]-U^2\left(1-2MU\right)\right\}\,.\nonumber\\
\end{eqnarray}
\end{small}
Equation (\ref{t}) is the non-commutative geodesic equation. If we
consider $MU<<1$, then the last equation simplifies to

\begin{equation}
U^{\prime\prime}+U=\frac{m^2M}{L^2}+3MU^2+\alpha^2\left[\frac{\left(E_0^2-m^2\right)U}{4L^2}-\frac{U^3}{4}\right]\,.\label{t2}
\end{equation}

In the reference \cite{PhysRevD.75.082001} is given a prescription
to calculate the deviation angle after one revolution. Let's recall
some ideas. First we consider a perturbation of Keplerian's
trajectory equation in the form
$$U^{\prime\prime}+U=\frac{M}{h^2}+\frac{f(U)}{h^2}\,,$$ where
$h^2=L^2/m^2$. In our case $\frac{f(U)}{h^2}=3MU^2+\alpha^2g(U)$.
Then it is possible to show that, after one revolution, there will
be an angle deviation given by
$$\Delta\phi=\frac{\pi f_1}{h^2}\,,$$  where $f_1=\frac{df(U)}{dU}\Big|_{U=1/l}$,  the distance $l$ is defined by
$l=a(1-e^2)$, with $a$ denoting the major semi-axis and $e$ the
eccentricity of the movement.

Hence if we compare our non-commutative geodesic equation to what is
given above, then we get the following angle deviation
\begin{equation}
\Delta\phi=\frac{6M\pi}{a(1-e^2)}+\alpha^2\pi\left[\frac{E_0^2/m^2-1}{4Ma(1-e^2)}-\frac{3}{4a^2(1-e^2)^2}\right]\,,
\label{desvio}
\end{equation}
it is possible to see that it comprises the well known general
relativity prediction and a correction which is given in terms of
the non-commutative parameter.

\subsection{Estimating the parameter $\alpha$}

In this subsection we will give a estimative of the parameter
$\alpha$ based on experimental measurements of $\Delta\phi$ for the
system Mercury-Sun. In such a system $m$ denotes the Mercury's mass
while $M$ stands for the Solar mass. Thus we will look for
experimental value for the precession of Mercury's perihelion to
compare with our expression (\ref{desvio}). In reference
\cite{lrr-2006-3} we find the following form for the precession
$$\Delta\phi_{exp}=\frac{6\pi M (1+\beta)}{a(1-e^2)}\,,$$ which is a way to determine deviations from what is predicted
by general relativity. Such a deviation is incorporated by the
parameter $\beta$, some experiments~\cite{lrr-2006-3} has bounded
its value as $\beta < 10^{-4}$. Therefore we find an expression for
$\alpha$ of the form

\begin{equation}
\alpha\simeq \frac{2M\sqrt{\beta}}{v}\,,
\end{equation}
where $v$ is
the planet's velocity, thus using orbital data of Mercury and the
experimental range of $\beta$ we finally find a range to settle the
non-commutative parameter, in standard units it reads

$$\alpha < 10^2 Km\,,$$ which is a small value when compared to
astronomical distances involved in the system Mercury-Sun. This is
consistent with the approximation used to obtain eq. (\ref{desvio}).

\section{Conclusion}

In this paper we have obtained the geodesic equation that arises
from the metric tensor which have been corrected using
non-commutative product between tetrads. Then we find corrections of
the deviation angle per revolution in terms of the non-commutative
parameter. Comparing our expression to experimental data it is
possible to estimate the range in which such parameter should lay.
We have find $\alpha\leq 10^2 Km\,,$ to put this result in context
we recall that the solar radius is approximatively $700000\, Km$,
thus the distance established by the non-commutative parameter would
lay inside the sun. However the Schwarzschild solution (which was
used to get our non-commutative geodesic equation) is only valid
outside the solar mass distribution. Then the parameter $\alpha$
could be felt outside the Sun by experiments such as the precession
of Mercury's perihelion. We hope that experimentalists would refine
the accuracy of the tests to detect deviations from general
relativity predictions in what concerns the precession of Mercury's
perihelion. Indeed the search for more accurate experiments would
help to decide if the spacetime has a non-commutative structure.


\begin{thebibliography}{10}%
\makeatletter
\providecommand \@ifxundefined [1]{%
 \ifx #1\undefined \expandafter \@firstoftwo
 \else \expandafter \@secondoftwo
\fi
}%
\providecommand \@ifnum [1]{%
 \ifnum #1\expandafter \@firstoftwo
 \else \expandafter \@secondoftwo
\fi
}%
\providecommand \enquote [1]{``#1''}%
\providecommand \bibnamefont  [1]{#1}%
\providecommand \bibfnamefont [1]{#1}%
\providecommand \citenamefont [1]{#1}%
\providecommand\href[0]{\@sanitize\@href}%
\providecommand\@href[1]{\endgroup\@@startlink{#1}\endgroup\@@href}%
\providecommand\@@href[1]{#1\@@endlink}%
\providecommand \@sanitize [0]{\begingroup\catcode`\&12\catcode`\#12\relax}%
\@ifxundefined \pdfoutput {\@firstoftwo}{%
 \@ifnum{\z@=\pdfoutput}{\@firstoftwo}{\@secondoftwo}%
}{%
 \providecommand\@@startlink[1]{\leavevmode}%
 \providecommand\@@endlink[0]{}%
}{%
 \providecommand\@@startlink[1]{%
  \leavevmode
  \pdfstartlink
   attr{/Border[0 0 1 ]/H/I/C[0 1 1]}%
   user{/Subtype/Link/A<</Type/Action/S/URI/URI(#1)>>}%
  \relax
 }%
 \providecommand\@@endlink[0]{\pdfendlink}%
}%
\providecommand \url  [0]{\begingroup\@sanitize \@url }%
\providecommand \@url [1]{\endgroup\@href {#1}{\urlprefix}}%
\providecommand \urlprefix [0]{URL }%
\providecommand \Eprint[0]{\href }%
\@ifxundefined \urlstyle {%
  \providecommand \doi [1]{doi:\discretionary{}{}{}#1}%
}{%
  \providecommand \doi [0]{doi:\discretionary{}{}{}\begingroup
  \urlstyle{rm}\Url }%
}%
\providecommand \doibase [0]{http://dx.doi.org/}%
\providecommand \Doi[1]{\href{\doibase#1}}%
\providecommand \bibAnnote [3]{%
  \BibitemShut{#1}%
  \begin{quotation}\noindent
    \textsc{Key:}\ #2\\\textsc{Annotation:}\ #3%
  \end{quotation}%
}%
\providecommand \bibAnnoteFile [2]{%
  \IfFileExists{#2}{\bibAnnote {#1} {#2} {\input{#2}}}{}%
}%
\providecommand \typeout [0]{\immediate \write \m@ne }%
\providecommand \selectlanguage [0]{\@gobble}%
\providecommand \bibinfo [0]{\@secondoftwo}%
\providecommand \bibfield [0]{\@secondoftwo}%
\providecommand \translation [1]{[#1]}%
\providecommand \BibitemOpen[0]{}%
\providecommand \bibitemStop [0]{}%
\providecommand \bibitemNoStop [0]{.\EOS\space}%
\providecommand \EOS [0]{\spacefactor3000\relax}%
\providecommand \BibitemShut [1]{\csname bibitem#1\endcsname}%
\bibitem{mp1}%
  \BibitemOpen
  \bibfield{author}{%
  \bibinfo {author} {\bibfnamefont{N.~T.}\ \bibnamefont{Roseveare}},\ }%
  \emph{\bibinfo {title} {Mercury's perihelion from Le Verrier to Einstein}},\
  Oxford science publications\ (\bibinfo {publisher} {Oxford University
  Press},\ \bibinfo {address} {USA},\ \bibinfo {year} {1982})%
  \bibAnnoteFile{NoStop}{mp1}%
\bibitem{mp2}%
  \BibitemOpen
  \bibfield{author}{%
  \bibinfo {author} {\bibfnamefont{A.}~\bibnamefont{Pais}},\ }%
  \emph{\bibinfo {title} {Subtle is the Lord. The Science and Life of Albert
  Einstein}}\ (\bibinfo {publisher} {Oxford University Press},\ \bibinfo
  {address} {USA},\ \bibinfo {year} {2005})%
  \bibAnnoteFile{NoStop}{mp2}%
\bibitem{mp3}%
  \BibitemOpen
  \bibfield{author}{%
  \bibinfo {author} {\bibfnamefont{S.}~\bibnamefont{Benczik}}, \bibinfo
  {author} {\bibfnamefont{L.~N.}\ \bibnamefont{Chang}}, \bibinfo {author}
  {\bibfnamefont{D.}~\bibnamefont{Minic}}, \bibinfo {author}
  {\bibfnamefont{N.}~\bibnamefont{Okamura}}, \bibinfo {author}
  {\bibfnamefont{S.}~\bibnamefont{Rayyan}},\ and\ \bibinfo {author}
  {\bibfnamefont{T.}~\bibnamefont{Takeuchi}},\ }%
  \bibfield{journal}{%
  \Doi{10.1103/PhysRevD.66.026003}{\bibinfo {journal} {Phys. Rev. D}}\ }%
  \textbf{\bibinfo {volume} {66}},\ \bibinfo {pages} {026003} (\bibinfo {month}
  {Jun}\ \bibinfo {year} {2002}),\
  \url{http://link.aps.org/doi/10.1103/PhysRevD.66.026003}%
  \bibAnnoteFile{NoStop}{mp3}%
\bibitem{mp4}%
  \BibitemOpen
  \bibfield{author}{%
  \bibinfo {author} {\bibfnamefont{B.}~\bibnamefont{Mirza}}\ and\ \bibinfo
  {author} {\bibfnamefont{M.}~\bibnamefont{Dehghani}},\ }%
  \bibfield{journal}{%
  \bibinfo {journal} {Commun. Theor. Phys.}\ }%
  \textbf{\bibinfo {volume} {42}},\ \bibinfo {pages} {183} (\bibinfo {year}
  {2004})%
  \bibAnnoteFile{NoStop}{mp4}%
\bibitem{mp5}%
  \BibitemOpen
  \bibfield{author}{%
  \bibinfo {author} {\bibfnamefont{J.~D.}\ \bibnamefont{Vergara}}\ and\
  \bibinfo {author} {\bibfnamefont{J.~M.}\ \bibnamefont{Romero}},\ }%
  \bibfield{journal}{%
  \Doi{10.1142/S0217732303011472}{\bibinfo {journal} {Modern Physics Letters
  A}}\ }%
  \textbf{\bibinfo {volume} {18}},\ \bibinfo {pages} {1673} (\bibinfo {year}
  {2003}),\
  \Eprint{http://arxiv.org/abs/http://www.worldscientific.com/doi/pdf/10.1142/%
S0217732303011472}{http://www.worldscientific.com/doi/pdf/10.1142/S02177323030%
11472},\
  \url{http://www.worldscientific.com/doi/abs/10.1142/S0217732303011472}%
  \bibAnnoteFile{NoStop}{mp5}%
\bibitem{mp6}%
  \BibitemOpen
  \bibfield{author}{%
  \bibinfo {author} {\bibfnamefont{K.}~\bibnamefont{Nozari}}\ and\ \bibinfo
  {author} {\bibfnamefont{S.}~\bibnamefont{Akhshabi}},\ }%
  \bibfield{journal}{%
  \Doi{10.1016/j.chaos.2006.09.042}{\bibinfo {journal} {Chaos Solitons
  Fractals}}\ }%
  \textbf{\bibinfo {volume} {37}},\ \bibinfo {pages} {324} (\bibinfo {year}
  {2008}),\ \Eprint{http://arxiv.org/abs/gr-qc/0608076}{arXiv:gr-qc/0608076
  [gr-qc]}%
  \bibAnnoteFile{NoStop}{mp6}%
\bibitem{weyl}%
  \BibitemOpen
  \bibfield{author}{%
  \bibinfo {author} {\bibfnamefont{H.}~\bibnamefont{Weyl}},\ }%
  \bibfield{journal}{%
  \Doi{10.1007/BF02055756}{\bibinfo {journal} {Zeitschrift für Physik}}\ }%
  \textbf{\bibinfo {volume} {46}},\ \bibinfo {pages} {1} (\bibinfo {year}
  {1927}),\ ISSN \bibinfo {issn} {0044-3328},\
  \url{http://dx.doi.org/10.1007/BF02055756}%
  \bibAnnoteFile{NoStop}{weyl}%
\bibitem{snyder1}%
  \BibitemOpen
  \bibfield{author}{%
  \bibinfo {author} {\bibfnamefont{H.~S.}\ \bibnamefont{Snyder}},\ }%
  \bibfield{journal}{%
  \Doi{10.1103/PhysRev.71.38}{\bibinfo {journal} {Phys.Rev.}}\ }%
  \textbf{\bibinfo {volume} {71}},\ \bibinfo {pages} {38} (\bibinfo {year}
  {1947})%
  \bibAnnoteFile{NoStop}{snyder1}%
\bibitem{snyder2}%
  \BibitemOpen
  \bibfield{author}{%
  \bibinfo {author} {\bibfnamefont{H.~S.}\ \bibnamefont{Snyder}},\ }%
  \bibfield{journal}{%
  \Doi{10.1103/PhysRev.72.68}{\bibinfo {journal} {Phys. Rev.}}\ }%
  \textbf{\bibinfo {volume} {72}},\ \bibinfo {pages} {68} (\bibinfo {month}
  {Jul}\ \bibinfo {year} {1947}),\
  \url{http://link.aps.org/doi/10.1103/PhysRev.72.68}%
  \bibAnnoteFile{NoStop}{snyder2}%
\bibitem{kalau}%
  \BibitemOpen
  \bibfield{author}{%
  \bibinfo {author} {\bibfnamefont{W.}~\bibnamefont{Kalau}}\ and\ \bibinfo
  {author} {\bibfnamefont{M.}~\bibnamefont{Walze}},\ }%
  \bibfield{journal}{%
  \Doi{10.1016/0393-0440(94)00032-Y}{\bibinfo {journal} {Journal of Geometry
  and Physics}}\ }%
  \textbf{\bibinfo {volume} {16}},\ \bibinfo {pages} {327 } (\bibinfo {year}
  {1995}),\ ISSN \bibinfo {issn} {0393-0440},\
  \url{http://www.sciencedirect.com/science/article/pii/039304409400032Y}%
  \bibAnnoteFile{NoStop}{kalau}%
\bibitem{kastler}%
  \BibitemOpen
  \bibfield{author}{%
  \bibinfo {author} {\bibfnamefont{D.}~\bibnamefont{Kastler}},\ }%
  \bibfield{journal}{%
  \bibinfo {journal} {Communications in Mathematical Physics}\ }%
  \textbf{\bibinfo {volume} {166}},\ \bibinfo {pages} {633} (\bibinfo {year}
  {1995}),\ ISSN \bibinfo {issn} {0010-3616},\ \bibinfo {note}
  {10.1007/BF02099890},\ \url{http://dx.doi.org/10.1007/BF02099890}%
  \bibAnnoteFile{NoStop}{kastler}%
\bibitem{connes1}%
  \BibitemOpen
  \bibfield{author}{%
  \bibinfo {author} {\bibfnamefont{A.~H.}\ \bibnamefont{Chamseddine}}\ and\
  \bibinfo {author} {\bibfnamefont{A.}~\bibnamefont{Connes}},\ }%
  \bibfield{journal}{%
  \bibinfo {journal} {Communications in Mathematical Physics}\ }%
  \textbf{\bibinfo {volume} {186}},\ \bibinfo {pages} {731} (\bibinfo {year}
  {1997}),\ ISSN \bibinfo {issn} {0010-3616},\ \bibinfo {note}
  {10.1007/s002200050126},\ \url{http://dx.doi.org/10.1007/s002200050126}%
  \bibAnnoteFile{NoStop}{connes1}%
\bibitem{connes2}%
  \BibitemOpen
  \bibfield{author}{%
  \bibinfo {author} {\bibfnamefont{A.}~\bibnamefont{Connes}}\ and\ \bibinfo
  {author} {\bibfnamefont{J.}~\bibnamefont{Lott}},\ }%
  \bibfield{journal}{%
  \Doi{10.1016/0920-5632(91)90120-4}{\bibinfo {journal} {Nuclear Physics B -
  Proceedings Supplements}}\ }%
  \textbf{\bibinfo {volume} {18}},\ \bibinfo {pages} {29 } (\bibinfo {year}
  {1991}),\ ISSN \bibinfo {issn} {0920-5632},\
  \url{http://www.sciencedirect.com/science/article/pii/0920563291901204}%
  \bibAnnoteFile{NoStop}{connes2}%
\bibitem{varilly1}%
  \BibitemOpen
  \bibfield{author}{%
  \bibinfo {author} {\bibfnamefont{J.~C.}\ \bibnamefont{VÃ¡rilly}}\ and\
  \bibinfo {author} {\bibfnamefont{J.}~\bibnamefont{Gracia-BondÃ­a}},\ }%
  \bibfield{journal}{%
  \Doi{10.1016/0393-0440(93)90038-G}{\bibinfo {journal} {Journal of Geometry
  and Physics}}\ }%
  \textbf{\bibinfo {volume} {12}},\ \bibinfo {pages} {223 } (\bibinfo {year}
  {1993}),\ ISSN \bibinfo {issn} {0393-0440},\
  \url{http://www.sciencedirect.com/science/article/pii/039304409390038G}%
  \bibAnnoteFile{NoStop}{varilly1}%
\bibitem{varilly2}%
  \BibitemOpen
  \bibfield{author}{%
  \bibinfo {author} {\bibfnamefont{C.}~\bibnamefont{MartÃ­n}}, \bibinfo
  {author} {\bibfnamefont{J.}~\bibnamefont{Gracia-BondÃ­a}},\ and\ \bibinfo
  {author} {\bibfnamefont{J.~C.}\ \bibnamefont{VÃ¡rilly}},\ }%
  \bibfield{journal}{%
  \Doi{10.1016/S0370-1573(97)00053-7}{\bibinfo {journal} {Physics Reports}}\ }%
  \textbf{\bibinfo {volume} {294}},\ \bibinfo {pages} {363 } (\bibinfo {year}
  {1998}),\ ISSN \bibinfo {issn} {0370-1573},\
  \url{http://www.sciencedirect.com/science/article/pii/S0370157397000537}%
  \bibAnnoteFile{NoStop}{varilly2}%
\bibitem{Szabo2003207}%
  \BibitemOpen
  \bibfield{author}{%
  \bibinfo {author} {\bibfnamefont{R.~J.}\ \bibnamefont{Szabo}},\ }%
  \bibfield{journal}{%
  \Doi{http://dx.doi.org/10.1016/S0370-1573(03)00059-0}{\bibinfo {journal}
  {Physics Reports}}\ }%
  \textbf{\bibinfo {volume} {378}},\ \bibinfo {pages} {207 } (\bibinfo {year}
  {2003}),\ ISSN \bibinfo {issn} {0370-1573},\
  \url{http://www.sciencedirect.com/science/article/pii/S0370157303000590}%
  \bibAnnoteFile{NoStop}{Szabo2003207}%
\bibitem{Seiberg:1999vs}%
  \BibitemOpen
  \bibfield{author}{%
  \bibinfo {author} {\bibfnamefont{N.}~\bibnamefont{Seiberg}}\ and\ \bibinfo
  {author} {\bibfnamefont{E.}~\bibnamefont{Witten}},\ }%
  \bibfield{journal}{%
  \bibinfo {journal} {JHEP}\ }%
  \textbf{\bibinfo {volume} {9909}},\ \bibinfo {pages} {032} (\bibinfo {year}
  {1999}),\ \Eprint{http://arxiv.org/abs/hep-th/9908142}{arXiv:hep-th/9908142
  [hep-th]}%
  \bibAnnoteFile{NoStop}{Seiberg:1999vs}%
\bibitem{belissard}%
  \BibitemOpen
  \bibfield{author}{%
  \bibinfo {author} {\bibfnamefont{J.}~\bibnamefont{{Bellissard}}}, \bibinfo
  {author} {\bibfnamefont{A.}~\bibnamefont{{van Elst}}},\ and\ \bibinfo
  {author} {\bibfnamefont{H.}~\bibnamefont{{Schulz-Baldes}}},\ }%
  \bibfield{journal}{%
  \Doi{10.1063/1.530758}{\bibinfo {journal} {Journal of Mathematical Physics}}\
  }%
  \textbf{\bibinfo {volume} {35}},\ \bibinfo {pages} {5373} (\bibinfo {month}
  {Oct.}\ \bibinfo {year} {1994}),\
  \Eprint{http://arxiv.org/abs/arXiv:cond-mat/9411052}{arXiv:cond-mat/9411052}%
  \bibAnnoteFile{NoStop}{belissard}%
\bibitem{PhysRevD.69.045013}%
  \BibitemOpen
  \bibfield{author}{%
  \bibinfo {author} {\bibfnamefont{R.}~\bibnamefont{Amorim}}\ and\ \bibinfo
  {author} {\bibfnamefont{F.~A.}\ \bibnamefont{Farias}},\ }%
  \bibfield{journal}{%
  \Doi{10.1103/PhysRevD.69.045013}{\bibinfo {journal} {Phys. Rev. D}}\ }%
  \textbf{\bibinfo {volume} {69}},\ \bibinfo {pages} {045013} (\bibinfo {month}
  {Feb}\ \bibinfo {year} {2004}),\
  \url{http://link.aps.org/doi/10.1103/PhysRevD.69.045013}%
  \bibAnnoteFile{NoStop}{PhysRevD.69.045013}%
\bibitem{Aschieri:2005yw}%
  \BibitemOpen
  \bibfield{author}{%
  \bibinfo {author} {\bibfnamefont{P.}~\bibnamefont{Aschieri}}, \bibinfo
  {author} {\bibfnamefont{C.}~\bibnamefont{Blohmann}}, \bibinfo {author}
  {\bibfnamefont{M.}~\bibnamefont{Dimitrijevic}}, \bibinfo {author}
  {\bibfnamefont{F.}~\bibnamefont{Meyer}}, \bibinfo {author}
  {\bibfnamefont{P.}~\bibnamefont{Schupp}}, \emph{et~al.},\ }%
  \bibfield{journal}{%
  \Doi{10.1088/0264-9381/22/17/011}{\bibinfo {journal} {Class.Quant.Grav.}}\ }%
  \textbf{\bibinfo {volume} {22}},\ \bibinfo {pages} {3511} (\bibinfo {year}
  {2005}),\ \Eprint{http://arxiv.org/abs/hep-th/0504183}{arXiv:hep-th/0504183
  [hep-th]}%
  \bibAnnoteFile{NoStop}{Aschieri:2005yw}%
\bibitem{UR}%
  \BibitemOpen
  \bibfield{author}{%
  \bibinfo {author} {\bibfnamefont{S.~C.}\ \bibnamefont{Ulhoa}}\ and\ \bibinfo
  {author} {\bibfnamefont{R.~G.~G.}\ \bibnamefont{Amorim}},\ }%
  \bibfield{journal}{%
  \bibinfo {journal} {Journal of Gravity}\ }%
  \textbf{\bibinfo {volume} {2013}},\ \bibinfo {pages} {217813} (\bibinfo
  {year} {2013})%
  \bibAnnoteFile{NoStop}{UR}%
\bibitem{Mukherjee:2006nd}%
  \BibitemOpen
  \bibfield{author}{%
  \bibinfo {author} {\bibfnamefont{P.}~\bibnamefont{Mukherjee}}\ and\ \bibinfo
  {author} {\bibfnamefont{A.}~\bibnamefont{Saha}},\ }%
  \bibfield{journal}{%
  \Doi{10.1103/PhysRevD.74.027702}{\bibinfo {journal} {Phys.Rev.}}\ }%
  \textbf{\bibinfo {volume} {D74}},\ \bibinfo {pages} {027702} (\bibinfo {year}
  {2006}),\ \Eprint{http://arxiv.org/abs/hep-th/0605287}{arXiv:hep-th/0605287
  [hep-th]}%
  \bibAnnoteFile{NoStop}{Mukherjee:2006nd}%
\bibitem{Chaichian2008573}%
  \BibitemOpen
  \bibfield{author}{%
  \bibinfo {author} {\bibfnamefont{M.}~\bibnamefont{Chaichian}}, \bibinfo
  {author} {\bibfnamefont{A.}~\bibnamefont{Tureanu}},\ and\ \bibinfo {author}
  {\bibfnamefont{G.}~\bibnamefont{Zet}},\ }%
  \bibfield{journal}{%
  \Doi{10.1016/j.physletb.2008.01.029}{\bibinfo {journal} {Physics Letters B}}\
  }%
  \textbf{\bibinfo {volume} {660}},\ \bibinfo {pages} {573 } (\bibinfo {year}
  {2008}),\ ISSN \bibinfo {issn} {0370-2693},\
  \url{http://www.sciencedirect.com/science/article/pii/S037026930800097X}%
  \bibAnnoteFile{NoStop}{Chaichian2008573}%
\bibitem{PhysRevD.75.082001}%
  \BibitemOpen
  \bibfield{author}{%
  \bibinfo {author} {\bibfnamefont{G.~S.}\ \bibnamefont{Adkins}}\ and\ \bibinfo
  {author} {\bibfnamefont{J.}~\bibnamefont{McDonnell}},\ }%
  \bibfield{journal}{%
  \Doi{10.1103/PhysRevD.75.082001}{\bibinfo {journal} {Phys. Rev. D}}\ }%
  \textbf{\bibinfo {volume} {75}},\ \bibinfo {pages} {082001} (\bibinfo {month}
  {Apr}\ \bibinfo {year} {2007}),\
  \url{http://link.aps.org/doi/10.1103/PhysRevD.75.082001}%
  \bibAnnoteFile{NoStop}{PhysRevD.75.082001}%
\bibitem{lrr-2006-3}%
  \BibitemOpen
  \bibfield{author}{%
  \bibinfo {author} {\bibfnamefont{C.~M.}\ \bibnamefont{Will}},\ }%
  \bibfield{journal}{%
  \bibinfo {journal} {Living Reviews in Relativity}\ }%
  \textbf{\bibinfo {volume} {9}} (\bibinfo {year} {2006}),\ \doi{\bibinfo {doi}
  {10.12942/lrr-2006-3}},\ \url{http://www.livingreviews.org/lrr-2006-3}%
  \bibAnnoteFile{NoStop}{lrr-2006-3}%
\end{thebibliography}

%

\end{document}